\documentclass{aastex631}

\shorttitle{Responses of a Coronal Hole to a Coronal Wave}
\shortauthors{Zhang et al.}
\graphicspath{{./}{figures/}}
\usepackage{soul}

\begin{document}

\title{Responses of a Coronal Hole to a Fast Flare-Driven Coronal Wave}


\correspondingauthor{Huadong Chen, Guiping Zhou}
\email{hdchen@nao.cas.cn, gpzhou@nao.cas.cn}

\author[0009-0005-2234-4319]{Xiaofan Zhang}
\affiliation{State Key Laboratory of Solar Activity and Space Weather, National Atronomical Observatories, Chinese Academy of Sciences, Beijing 100101, China}
\affiliation{State Key Laboratory of Solar Activity and Space Weather, National Space Science Center, Chinese Academy of Sciences, Beijing 100190, China}
\affiliation{University of Chinese Academy of Sciences, Beijing 100049, China}

\author[0000-0001-6076-9370]{Huadong Chen}
\affiliation{State Key Laboratory of Solar Activity and Space Weather, National Atronomical Observatories, Chinese Academy of Sciences, Beijing 100101, China}
\affiliation{State Key Laboratory of Solar Activity and Space Weather, National Space Science Center, Chinese Academy of Sciences, Beijing 100190, China}
\affiliation{University of Chinese Academy of Sciences, Beijing 100049, China}

\author[0000-0001-8228-565X]{Guiping Zhou}
\affiliation{State Key Laboratory of Solar Activity and Space Weather, National Atronomical Observatories, Chinese Academy of Sciences, Beijing 100101, China}
\affiliation{State Key Laboratory of Solar Activity and Space Weather, National Space Science Center, Chinese Academy of Sciences, Beijing 100190, China}
\affiliation{University of Chinese Academy of Sciences, Beijing 100049, China}

\author[0000-0003-4655-6939]{Li Feng}
\affiliation{Key Laboratory of Dark Matter and Space Astronomy, Purple Mountain Observatory, Chinese Academy of Sciences, Nanjing 210023, China}

\author[0000-0002-4241-9921]{Yang Su}
\affiliation{Key Laboratory of Dark Matter and Space Astronomy, Purple Mountain Observatory, Chinese Academy of Sciences, Nanjing 210023, China}

\author[0000-0002-4205-5566]{Jinhan Guo}
\affiliation{School of Astronomy and Space Science, Nanjing
  University, Nanjing 210023, China}

\author[0000-0001-5776-056X]{Leping Li}
\affiliation{State Key Laboratory of Solar Activity and Space Weather, National Atronomical Observatories, Chinese Academy of Sciences, Beijing 100101, China}
\affiliation{State Key Laboratory of Solar Activity and Space Weather, National Space Science Center, Chinese Academy of Sciences, Beijing 100190, China}
\affiliation{University of Chinese Academy of Sciences, Beijing 100049, China}

\author{Wei Lin}
\affiliation{State Key Laboratory of Solar Activity and Space Weather, National Atronomical Observatories, Chinese Academy of Sciences, Beijing 100101, China}
\affiliation{State Key Laboratory of Solar Activity and Space Weather, National Space Science Center, Chinese Academy of Sciences, Beijing 100190, China}
\affiliation{University of Chinese Academy of Sciences, Beijing 100049, China}

\author[0000-0002-5431-6065]{Suli Ma}
\affiliation{State Key Laboratory of Solar Activity and Space Weather, National Space Science Center, Chinese Academy of Sciences, Beijing 100190, China}
\affiliation{University of Chinese Academy of Sciences, Beijing 100049, China}

\author[0000-0001-9493-4418]{Yuandeng Shen}
\affiliation{State Key Laboratory of Solar Activity and Space Weather, School of Aerospace, Harbin Institute of Technology, Shenzhen 518055, China}
\affiliation{Shenzhen Key Laboratory of Numerical Prediction for Space Storm, Harbin Institute of Technology, Shenzhen 518055, China}

\author[0000-0002-2734-8969]{Ruisheng Zheng}
\affiliation{Shandong Provincial Key Laboratory of Optical Astronomy and Solar-Terrestrial Environment, and Institute of Space Sciences, Shandong University, Weihai 264209, China}

\author[0000-0002-1396-7603]{Suo Liu}
\affiliation{State Key Laboratory of Solar Activity and Space Weather, National Atronomical Observatories, Chinese Academy of Sciences, Beijing 100101, China}
\affiliation{State Key Laboratory of Solar Activity and Space Weather, National Space Science Center, Chinese Academy of Sciences, Beijing 100190, China}
\affiliation{University of Chinese Academy of Sciences, Beijing 100049, China}

\author[0000-0003-2686-9153]{Xianyong Bai}
\affiliation{State Key Laboratory of Solar Activity and Space Weather, National Atronomical Observatories, Chinese Academy of Sciences, Beijing 100101, China}
\affiliation{State Key Laboratory of Solar Activity and Space Weather, National Space Science Center, Chinese Academy of Sciences, Beijing 100190, China}
\affiliation{University of Chinese Academy of Sciences, Beijing 100049, China}

\author{Yuanyong Deng}
\affiliation{State Key Laboratory of Solar Activity and Space Weather, National Atronomical Observatories, Chinese Academy of Sciences, Beijing 100101, China}
\affiliation{State Key Laboratory of Solar Activity and Space Weather, National Space Science Center, Chinese Academy of Sciences, Beijing 100190, China}
\affiliation{University of Chinese Academy of Sciences, Beijing 100049, China}

\author{Jingxiu Wang}
\affiliation{University of Chinese Academy of Sciences, Beijing 100049, China}
\affiliation{State Key Laboratory of Solar Activity and Space Weather, National Atronomical Observatories, Chinese Academy of Sciences, Beijing 100101, China}
\affiliation{State Key Laboratory of Solar Activity and Space Weather, National Space Science Center, Chinese Academy of Sciences, Beijing 100190, China}



\begin{abstract}

Coronal waves, significant solar phenomena, act as diagnostic tools for scientists studying solar atmosphere properties. Here, we present a novel observation detailing how a coronal wave event, associated with an X5.0 class flare, influenced the properties of an adjacent coronal hole through interaction. The coronal wave was observed in both extreme ultraviolet observations from the Atmospheric Imaging Assembly aboard the Solar Dynamics Observatory and Lyman-alpha observations from the Solar Disk Imager aboard the Advanced Space-based Solar Observatory. Utilizing the method of differential emission measure, we found that as the coronal wave passed through, the adjacent coronal hole experienced an increase in temperature from 1.31 to 1.43 MK and a rise in density from $\sim$1.62$\times10^{8}$ to 1.76$\times10^{8}$ cm$^{-3}$ within the rising period of $\sim$7 minutes. Subsequently, after the wave passed, the entire coronal hole transitioned to a new state with a slight temperature increase and a 14$\%$ decrease in density, with more pronounced changes observed at the coronal hole's boundary. Taking into account the impacts of radiative loss and heat conduction, the coronal wave was estimated to provide an average energy of 2.2$\times10^{8}$ erg cm$^{-2}$ to the coronal hole during the short rising period. This study highlights the identification of the coronal wave in both extreme ultraviolet and Lyman-alpha observations, shedding light on the significant energy input, particularly within the coronal hole. These findings provide new insights into better understanding kinematics of fast coronal waves, energy transfer processes open versus closed magnetic topologies, and the possible acceleration of solar winds.

\end{abstract}


\section{Introduction} \label{sec:intro}

Coronal waves, also known as large-scale propagating extreme ultraviolet (EUV) disturbances, are critical phenomena closely associated with eruptive events such as solar flares \citep{fletcher2011observational, wang2021exploring}, coronal mass ejections \citep[CMEs;][]{chen2011coronal, hou2022three}, and other injections \citep{warmuth2015large, zheng2024recent}. Observations of coronal waves in EUV, soft X-ray (SXR) emissions, and H$\alpha$ and \ion{He}{1}, have enabled detailed studies of their kinematics and energy transport mechanisms \citep{{2018ApJ...868..107V, 2014SoPh..289.3233L, 2018ApJ...864L..24L}}, and insights into the underlying processes of solar activity and dynamics.

The interaction between coronal waves and coronal holes (CHs) provides insights into energy transfer processes in open versus closed magnetic topologies. CHs, regions of open magnetic fields and reduced plasma density, are key sources of fast solar wind\citep{withbroe1977mass} that significantly influence the geospace environment. Previous studies have reported wave reflections\citep[e.g.][]{long2008kinematics, veronig2008high}, oscillations \citep{liu2012quasi}, refractions, or amplitude dampening at CH boundaries \citep{2011ApJ...727L..43K}. 
\citet{2009ApJ...691L.123G} observe the measurable reflection of EUV wave from a CH, noting that reflected waves were generally slower than the direct wave.
\citet{2022A&A...659A.164Z} report the total reflection of a flare-driven quasi-periodic EUV wave train at a CH boundary. Coronal wave can transmit into CHs \citep[][]{olmedo2012secondary}. When coronal wave traverse a uniquely shaped CH, their wavefronts can converge together at some specific point \citep[e.g.][]{zhou2024resolved}. Numerical studies reveal that the interaction between propagating waves and CHs generates reflected, traversing, trasmitted waves, as well as stationary features at one-sided CH boundary  \citep{2010ApJ...713.1008S, 2018ApJ...860...24P}.  Different CH geometries are showed to have effects on the density profiles of reflected waves in simulations \citep{2024A&A...687A.200P}.  In  an off-limb solar corona, EUV wave was suggested to increase the electron density by 12\%-18\% \citep[e.g.][]{2011ApJ...733L..25K}. \citet{2015ApJ...812..173V} proposed 6\%–9\% density and 5\%–6\% temperature increases during the wave passage of the local coronal plasma. However, quantitative analyses of thermodynamic changes within CHs induced by such interactions remain sparse, with most studies focusing on wave kinematics \citep[e.g.][]{2020ApJ...905..150Z} rather than plasma parameter evolution.

Recent advancements in differential emission measure (DEM) techniques have improved the capability to track temperature and density variations during transient events \citep{cheung2015thermal, 2012A&A...539A.146H}. Meanwhile, the coupling of EUV observations with Lyman-alpha diagnostics such as those from the Solar Disk Imager (SDI) aboard the Advanced Space-based Solar Observatory \citep[ASO-S;][]{gan2019advanced} offers new opportunities to trace wave-related disturbances across atmospheric layers \citep{2013ApJ...776...58N}. Despite these advancements, the energy budget transferred from coronal waves to CHs and its spatiotemporal distribution remain poorly constrained.

In this study, we present the first combined analysis of an X5.0 flare-driven coronal wave and its interaction with an adjacent CH using multi-instrument observations (SDO/AIA and ASO-S/SDI). We quantify transient and persistent changes in CH temperature, density, and energy balance, providing critical observational constraints on wave-CH energy transfer mechanisms. The data and observational results are shown in Section \ref{sec:obs}. Discussions and conclusions can be found in Section \ref{sec:dis}.

\section{Observations and Results} \label{sec:obs}

An X5.0 solar flare originating from active region (AR) NOAA 13536 began at 21:36 UT and reached its peak at 21:55 UT on 2023 December 31 \citep[e.g.][]{ryan2024triangulation}. This eruption and its associated coronal wave were not only captured by observations from the Atmospheric Imaging Assembly \citep[AIA;][]{lemen2012atmospheric} aboard the Solar Dynamics Observatory \citep[SDO;][]{pesnell2012solar}, but also by SDI of the Lyman-alpha Solar Telescope \citep[LST;][]{li2019lyman, chen2019lyman, feng2019lyman}, and the Hard X-ray Imager \citep[HXI;][]{zhang2019hard, su2019simulations} on board ASO-S. The EUV data from AIA featured a spatial resolution of 0$\farcs$6 per pixel and a cadence of 12 seconds, whereas the SDI Lyman-alpha full-disk data had resolutions of 1$\farcs$2 and 1 minute. The HXI observations had resolutions of $\sim$6$\farcs$5 and a 30-second cadence. The magnetic structures in the surrounding environment were examined using longitudinal magnetograms with a resolution of 0$\farcs$5 from the Helioseismic and Magnetic Imager \citep[HMI;][]{scherrer2012helioseismic} on board SDO.  Furthermore, observations from the H$\alpha$ Imaging Spectrograph \citep[HIS;][]{2022SCPMA..6589605L, 2022SCPMA..6589603Q} on the Chinese H$\alpha$ Solar Explorer \citep[CHASE;][]{2019RAA....19..165L, 2022SCPMA..6589602L} revealed a long filament near AR13536.

\begin{figure}
\centering
\includegraphics[width=0.9\textwidth, angle =0 ]{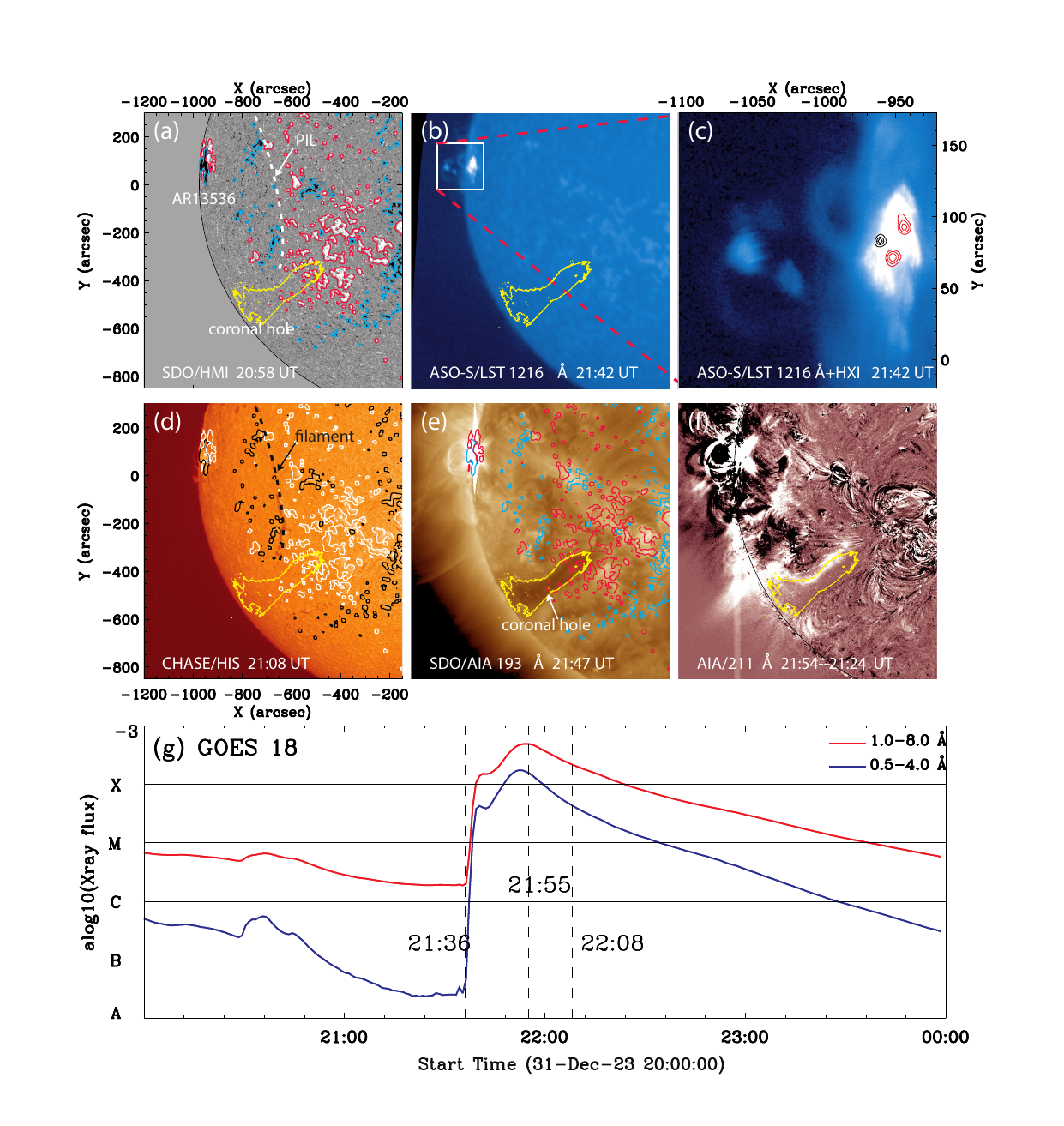}
\caption{Multi-wavelength observations revealing the flare-associated coronal wave and its sweeping region. Panel (a) displays an HMI image at 20:58 UT on December 31, showcasing AR13536 in N05E75, a nearby polarity inversion line (PIL) indicated by a white dashed line, and the examined CH with predominantly positive polarity outlined by yellow contours. In Panel (b), an image at 1216 \AA~from the LST illustrates the flare progression observed in Lyman-alpha. Panel (c) presents a subset of Panel (b) enclosed in a white box, superimposed with Hard X-ray contours representing energy ranges of 15-20 keV and 30-50 keV, respectively. Within the Chase/HIS observations depicted in Panel (d), a long filament (indicated by a black dashed line) runs along the PIL. Panel (e) exhibits an EUV image at 193 \AA~from AIA, revealing the flare activity in AR13536 and outlining the CH with yellow contours. The wave potentially accumulates at the CH boundary during its propagation, as evidenced by the difference image between 21:54-21:24 UT in 211 \AA~. Panel (g) showcases the X5 flaring process captured through GOES soft X-ray flux. The contours in Panels (a, d, and e) represent positive (red/white) and negative flux densities (cyan/black) at $\pm$15 G, respectively.} \label{fig:fig1}
\end{figure}

The notably intense solar eruption on 2023 December 31, featuring an X5-class flare and an associated halo coronal mass ejection (CME) with a speed of 2852 km s$^{-1}$ (https://cdaw.gsfc.nasa.gov). A rapid coronal wave was observed in both AIA/EUV and LST/UV observations. This significant eruption originated from AR13536 on the northeast limb. The coronal wave propagated and swept over the nearby PIL and CH, as depicted in Figure \ref{fig:fig1}(a). Lyman-alpha observations at 1216 \AA~from LST (refer to Figure \ref{fig:fig1}(b-c)) revealed the ejecting structure moving away from the east limb. Throughout the eruption, three sources of initial hard X-ray (HXR) emissions emerged: one at the top (indicated by black contours at 15-20 keV) and two at the footpoints (shown by red contours at 30-50 keV) of the associated post-flare loops (refer to Figure \ref{fig:fig1}(c)) \citep{su2013imaging} within AR13536. A long filament appeared along the long PIL in Figure \ref{fig:fig1}(d), indicating a magnetic interface. The CH was situated approximately 450 Mm southwest of AR13536, as illustrated in the AIA 193 \AA~observation in Figure \ref{fig:fig1}(e). The boundaries of the CH (indicated by the yellow curves) represent another type of magnetic interface. This fast coronal wave may elicit varied responses as it encounters different boundaries. This study focuses on exploring how the fast coronal wave influenced the physical properties of the CH.

\begin{figure}
\centering
\includegraphics[width=0.9\textwidth, angle =0 ]{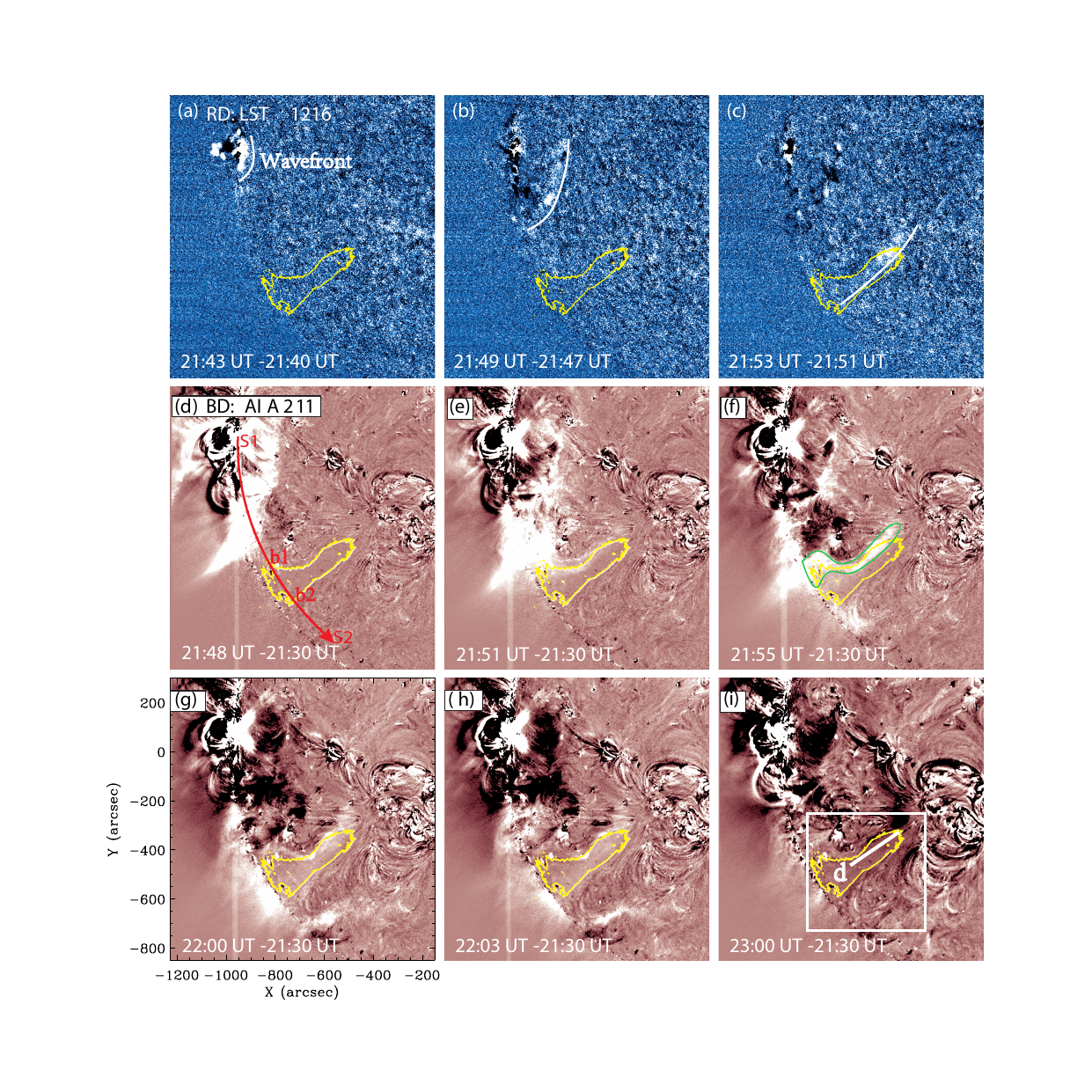}
\caption{The coronal wave is evident in the running difference UV images at 1216 \AA~ (top row) and the base difference EUV images at 211 \AA~(bottom two rows). The advancing fronts of the wave are delineated by white curves in Panels (a-b) and appear as white features in the bottom two rows. Notably, the wave exhibited an accumulation at the boundary of the CH, as indicated by the green contour in Panel (f). The CH is outlined by yellow curves in each panel. The white box in Panel (i) denotes the field of view of Figure \ref{fig:fig4}. The slice of ``S1-S2" in Panel (d) was chosen to generate the time-distance maps depicted in Figure \ref{fig:fig3}. The white bar ``d" in Panel (i) denoted the distance of $\sim$175Mm between the weight center of the CH and the right-side boundary. An animation of the AIA 211 \AA~base difference images (as shown in panels (d-i)) is available. The sequence spans from 21:30 UT to 22:21 UT on 2023 December 31,  with all frames computed relative to the initial 21:30 UT intensity image, illustrating the coronal wave propagation in EUV.	\label{fig:fig2}}
\end{figure}

\subsection{Propagation and Kinetics of the UV and EUV Wave}\label{sec:wave}

The coronal wave propagated through the CH was not only captured by the AIA/EUV observations but also by the LST/UV data, as illustrated in Figure \ref{fig:fig2}. Zhou et al. (2025, in preparation) were the pioneers in documenting such a coronal wave in the Lyman-alpha line at 1216 \AA, revealing specific insights from the corona. This Lyman-alpha wave demonstrates a notable velocity of up to approximately 1000 km s$^{-1}$ and seemed to traverse the CH.

In the EUV observations depicted in the lower rows of Figure \ref{fig:fig2}, this coronal wave was observed propagating and displaying a noticeable concentration at the boundary of the CH, as indicated by the green contours in Figure \ref{fig:fig2}(f). This concentration was interpreted as a form of stationary-mode wave transformed by the slow component of the coronal wave \citep[e.g.,][]{chen2016can, zheng2018extreme}. Concurrently with the coronal wave traversing the CH, it induced a subtle increase in EUV intensity within the CH, as shown in Figure \ref{fig:fig2}(g-h).

\begin{figure}
\centering
\includegraphics[width=0.9\textwidth, angle =0 ]{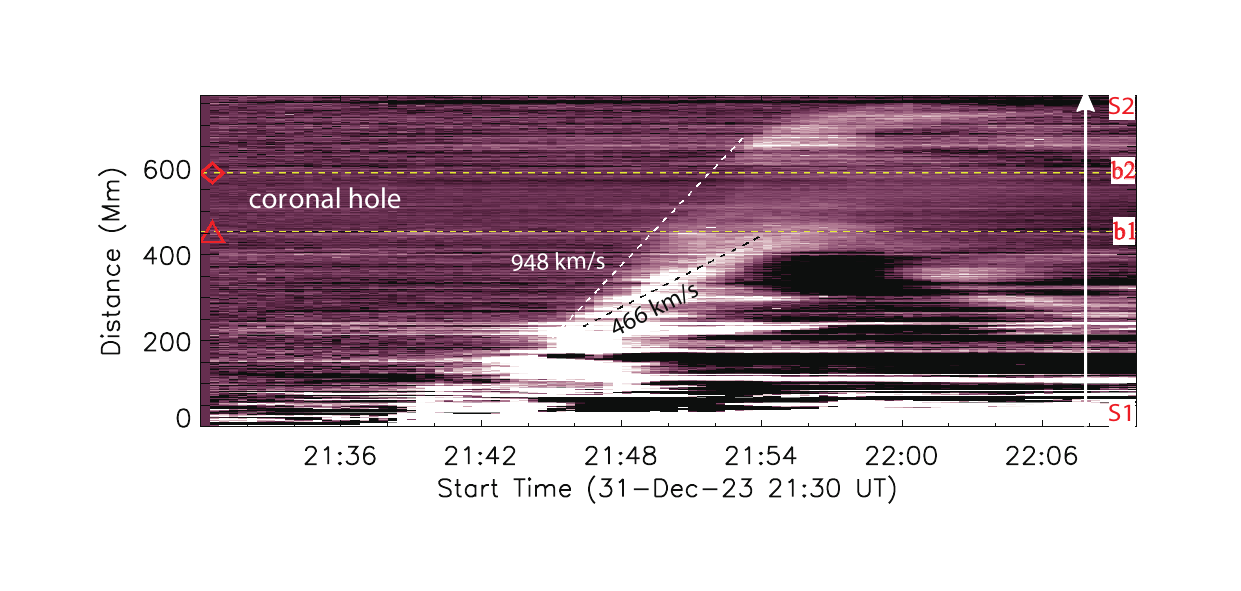}
\caption{Two types of coronal waves illustrated in the time-distance maps along the slice ``S1-S2'' as depicted in Figure \ref{fig:fig2}(d). The slice crosses the boundary of the CH twice, marked as ``b1'' and ``b2''. These two modes of coronal waves propagated over the CH, with velocities of 948 km s$^{-1}$ and 466 km s$^{-1}$ based on linear fitting. The positions of the CH boundary are indicated by two yellow horizontal dashed lines. \label{fig:fig3}}
\end{figure}

To illustrate how coronal waves transverse the CH, a time-distance diagram was created using the differential data at AIA 211 \AA~along the slice ``s1-s2'' as indicated in Figure \ref{fig:fig2}(d). The resulting diagrams are presented in Figure \ref{fig:fig3}. Concurrent with the onset of the flare/CME, we observed that the associated coronal waves exhibited two components: one moving rapidly at 948 km s$^{-1}$, and the other moving more slowly at 466 km s$^{-1}$ according to linear fitting. Given that the fast wave's speed is close to the typical Alfv\'{e}n speed in corona ($\sim$1000 km s$^{-1}$), this wave component is likely to be shocked \citep{2018ApJ...864L..24L}. The slow wave's speed far exceeds the acoustic speed of $\sim$230 km s$^{-1}$ in the quiet corona \citep{priest2014magnetohydrodynamics}, we suppose that it may be a fast-mode magnetosonic wave \citep{2012ApJ...750..134D} and/or the lateral front of the fast CME expansion \citep{chen2002evidence}. Enhanced EUV intensity corresponding to the waves indicated that the faster component could traverse the CH, whereas majority of the slower component halted at the CH boundary. In Figure \ref{fig:fig3}, two yellow horizontal dashed lines mark the boundaries of the CH. These observations suggest that the slower component might transform into a stationary-mode wave at the CH boundary, evident through the accumulation of EUV intensity there. The rapid component of the coronal waves exhibited a velocity similar to that of the Lyman-alpha wave at 1216 \AA ~ as depicted in Figure \ref{fig:fig2}(a-c). It would be intriguing to investigate how the CH responded to the passage of the swift wave in subsequent observations.

\subsection{Response of the CH to the Fast Coronal Wave}\label{sec:ch}

\begin{figure}
\centering
\includegraphics[width=1.\textwidth, angle =0 ]{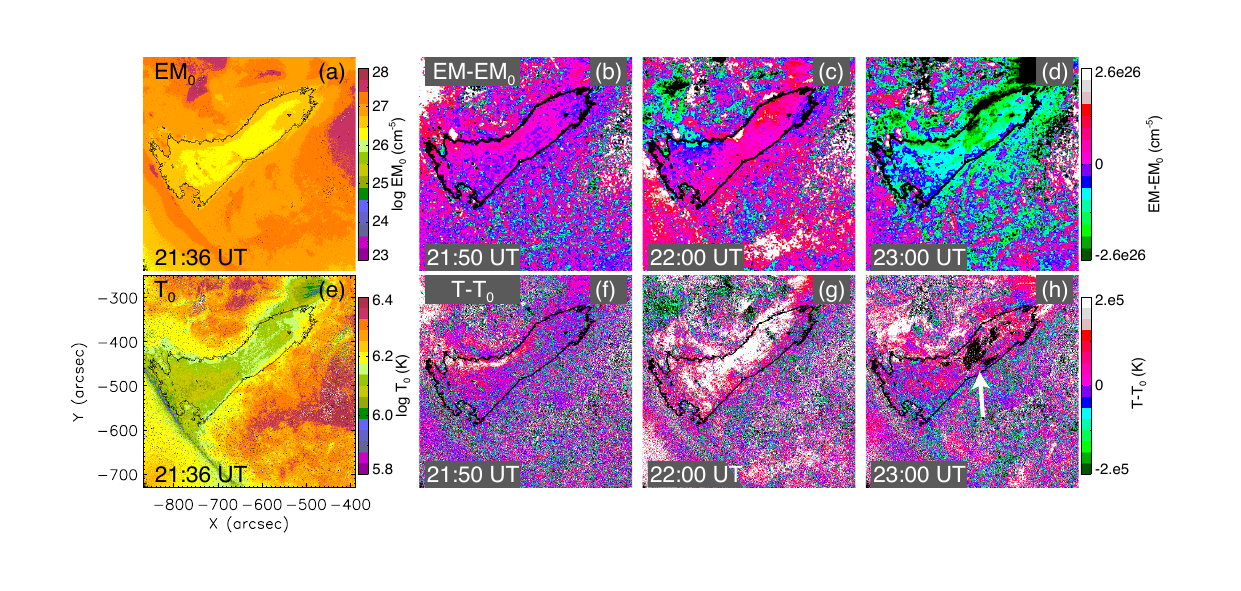}
\caption{Evolution and distribution of emission measure (EM) and temperature (T) within the CH delineated by black contours. The EM and T within the CH were estimated by DEM analysis using AIA data from six wavelengths: 131, 171, 193, 21, 335, and 94 \AA. Panel (a) and (e) display the EM and T distributions in the region before being perturbed by the fast coronal wave at 21:36 UT, labeled as $EM_0$ and $T_0$. Panels (b-d) depict the distribution of the base-difference EM obtained by subtracting $EM_0$, while similarly the T distributions in Panels (f-h) by subtracting $T_0$. In Panel (h), a white arrow indicates the black unavailable region with erroneous data points.  An animation of the EM (panels (b-d)) and T (panels (f-h)) base difference images is available. The sequence spans from 21:36 UT to 23:59 UT on 2023 December 31,  with all frames computed relative to the initial EM and T map at 21:36 UT, illustrating the changes of EM and T due to the interaction between the wave and the CH. \label{fig:fig4}}
\end{figure}

Using DEM analysis method \citep{cheung2015thermal}, we track the CH's EM and temperature evolution during wave passage (Figure \ref{fig:fig4}). Here, $T$ represents the DEM-weighted mean temperature, calculated as $\bar{T}=\frac{\int DEM(T) \times T dT}{\int DEM(T) dT}$ over the log T range [5.4, 6.9] \citep[e.g.,][]{chen2024simultaneous}. Initially, the CH exhibits lower EM and temperature values compared to those of the surrounding corona, as depicted in Figure \ref{fig:fig4}(a) and \ref{fig:fig4}(e), with average values of $\sim$ 3.2 $\times$ 10$^{26}$ cm$^{-5}$ and 1.31$\pm$0.02 MK, respectively.

Following the passage of the fast coronal wave across the CH starting at 21:50 UT, EM noticeably increased in the CH (Figure \ref{fig:fig4}(b-c)). However, approximately one hour after the fast coronal wave exited the CH, EM clearly decreased (Figure \ref{fig:fig4}(d)), dropping even below the levels observed in the CH before the disturbance. Regarding the temperature ($T$) in the CH, there was a significant rise following the transit of the fast coronal wave, maintaining a subtle enhancement even an hour after the wave had dissipated (Figure \ref{fig:fig4}(f-h)).

To accurately quantify the variations in plasma properties within the CH, we assessed the electron number density($n_e$) and energy. By utilizing the formula $n_e=\sqrt{EM/l}$, where $l$ represents the average line-of-sight (LOS) depth of the CH (considered as 123 Mm due to the sharp decrease in EUV intensity above this height over the east of the CH), we computed $n_e$ for the entire CH.

Subsequent to the passage of the fast coronal wave, the EM, T, and $n_e$ in the CH peaked within approximately 6-8 minutes and then declined to their minimum values after around 30 minutes. Based on these three parameters, we calculated the energy budget transferred from the waves to the CH and examined how this energy dissipates.

\subsection{Estimating Energy Input into the CH by the Fast Wave}\label{sec:en}

\begin{figure}
\centering
\includegraphics[width=0.8\textwidth, angle =0 ]{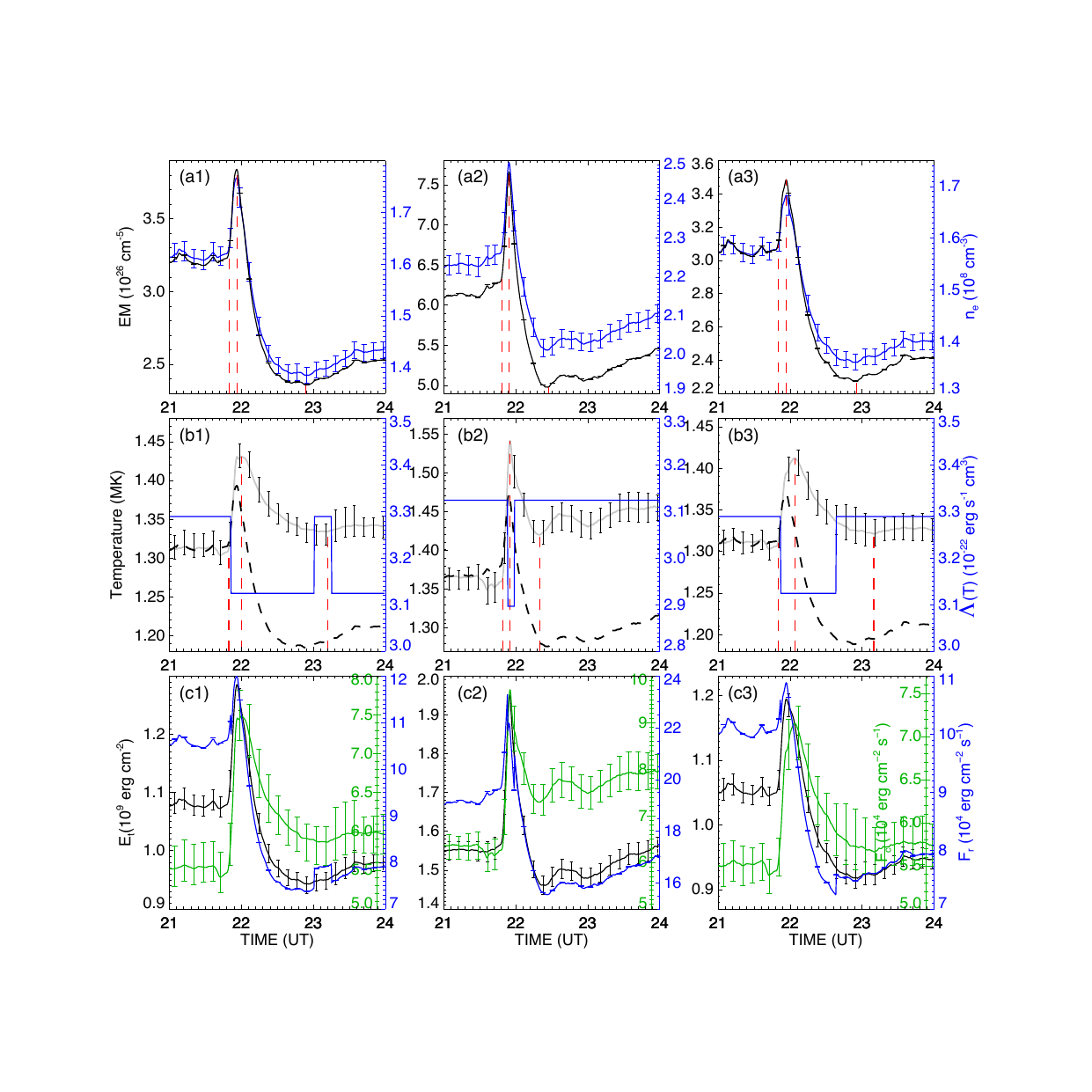}
\caption{Time profiles of essential parameters were analyzed to understand the dynamics within the CH. In Panels (a1-a2), the evolution of EM (black curves),  electron number density ($n_e$, blue curves) with 1$\sigma$ uncertainties are depicted. Panels (b1-b2) display the DEM-weighted mean temperature ($T$, gray curves) with 1$\sigma$ uncertainties, the adiabatic temperature (black dash curves) and radiative loss rates $\Lambda(T)$ (blue curves). Moving to Panels (c1-c2), thermal energy per unit area ($E_t$, black curves), radiative loss flux ($F_r$, blue curves), and heat conduction flux ($F_c$, green curves) with 1$\sigma$ uncertainties are illustrated. These profiles are generated for three distinct regions: the CH area (Panels (a2-c2)), the northern boundary of the CH (Panels (a3-c3)), and the primary part of the CH excluding the northern boundary. In the top two rows, three vertical dashed lines in each panel signify the onset, peak, and minimum decay points for the varying EM and T. The x-axis represents time, while the y-axes display the values of the respective parameters. \label{fig:fig5}}
\end{figure}

To comprehensively understand how the CH reacted to the fast coronal wave, we analyzed energy transmission in three specific regions: the CH area, the northern boundary marked by green curves in Figure \ref{fig:fig2}(f), and the main body of the CH excluding the northern boundary. This method was employed to minimize the potential impact of stationary-mode waves forming at the CH boundary \citep[e.g.,][]{chen2016can}. Additionally, erroneous data points, like those within the black area indicated by the white arrow in Figure \ref{fig:fig4}(h), were excluded from the calculations. Subsequently, four parameters were calculated (Figure \ref{fig:fig5}): the average thermal energy ($E_t$), the average radiative energy loss flux ($F_r$), the average heat conduction loss flux ($F_c$), and the kinetic energy flux ($F_k$).

Figure \ref{fig:fig5}(a1-a3) illustrate the temporal evolution of the EM and $n_e$ from 21:00 to 24:00 UT. Following the passage of the fast coronal wave, EM within the CH decreased by 25\%, dropping from $\sim$$3.2\times10^{26}$ cm$^{-5}$ to a minimum of $\sim$$2.4\times10^{26}$ cm$^{-5}$ within one hour. Regarding $n_e$ (blue curves in Figure \ref{fig:fig5}(a1-a3)), it exhibited a similar rapid increase and subsequent decline during the fast wave transit. Prior to the event, $n_e$ was measured at 1.62$\pm$0.02$\times10^8$ cm$^{-3}$ in the CH, akin to measurements in a south polar CH as observed in \cite{gallagher1999radial}. This value rose significantly by 38\% to about 2.23$\pm$0.02$\times10^8$ cm$^{-3}$ at the CH boundary and decreased by 3\% to approximately 1.57$\pm$0.02$\times10^8$ cm$^{-3}$ within the CH excluding its northern boundary. $n_e$ experienced a rapid initial increase followed by a subsequent decrease of approximately 14\% compared to its pre-disturbance level within the CH over the span of one hour.

The temporal evolution of the DEM-weighted mean temperature ($T$) and radiative loss rates $\Lambda(T)$ are depicted as gray and blue curves for the three distinct regions in Figure \ref{fig:fig5}(b1-b3). $\Lambda(T)$ represents the radiative loss rate based on the CHIANTI 8.0 database \citep{del2021chianti}. $T$ displays a rapid ascent and a gradual descent following the passage of the fast wave. Notably, a pronounced temperature increase at the CH boundary in Figure \ref{fig:fig5}(b2) was attributed to the influence of a stationary-mode wave. Within the CH, $T$ initially increased by approximately 9\% from 1.31$\pm$0.02 MK to 1.43$\pm$0.02 MK and then decreased to 1.34$\pm$0.02 MK about 1 hour later.

Given the parameters outlined earlier, we can calculate the energy input from the fast coronal wave into the CH. Assuming full ionization, the average thermal energy per unit area can be estimated using the formula $E_t = \frac{3}{2}2n_ekTl$ \citep{zhang2019energy}, where $k$ represents the Boltzmann constant and $l$ denotes the CH's line of sight (LOS) depth. Initially, $E_t$ increased from 1.08$\pm$0.02$\times10^9$ erg cm$^{-2}$ within the CH to a peak of 1.29$\pm$0.02$\times10^9$ erg cm$^{-2}$ before declining to 0.94$\pm$0.02$\times10^9$ erg cm$^{-2}$ (black curve in Figure \ref{fig:fig5}(c1)). The total thermal energy change $\Delta E_t$ was calculated as 2.1$\pm$0.3$\times10^8$ erg cm$^{-2}$ over the ascending phase.

During the rapid wave passage through the CH, extra energy dissipation may occur within the CH through mechanisms such as radiation or conduction. The average energy flux associated with radiative losses is determined by the formula $F_r = n_e^2\Lambda(T)l = EM\Lambda(T)$ \citep{priest2014magnetohydrodynamics}. As illustrated by the blue curves in Figure \ref{fig:fig5}(c1-c3), $F_r$ within the CH initially stood at $\sim$ 10.6$\times10^4$ erg s$^{-1}$cm$^{-2}$, peaked at $\sim$ 12.0$\times10^4$ erg s$^{-1}$cm$^{-2}$, and eventually decreased to $\sim$ 7.4$\times10^4$ erg s$^{-1}$cm$^{-2}$. By integrating over the rising period, the total increment of radiative energy loss $\Delta E_r$ within the CH was estimated to be $\sim$ 3.6$\times10^6$ erg cm$^{-2}$.

Another significant mode of energy dissipation is thermal conduction. Assuming force-free fields, the average flux of thermal conduction loss can be estimated as $F_c = -\kappa T^{5/2}\frac{dT}{ds} \approx -\frac{2}{7}\kappa\frac{T^{7/2}}{l}$, where the Spitzer conductivity is given by $\kappa = 9.2 \times 10^{-7}$ erg s$^{-1}$ cm$^{-1}$ K$^{-7/2}$ \citep[e.g.,][]{rosner1977hydrostatic, aschwanden2004role}. The thermal conduction flux $F_c$ within the CH increased from an initial value of 5.5$\pm$0.3$\times10^4$ erg s$^{-1}$ cm$^{-2}$ \citep[e.g.,][]{munro1972properties, huber1974extreme, withbroe1977mass}, peaked at 7.5$\pm$0.3$\times10^4$ erg s$^{-1}$ cm$^{-2}$, and eventually decreased to 5.9$\pm$0.3$\times10^4$ erg s$^{-1}$ cm$^{-2}$ (as indicated by the green curve in Figure \ref{fig:fig5}(c1)). The total increase of energy loss due to thermal conduction, denoted as $\Delta E_c$, within the CH over the rising period was estimated to be approximately 4.2$\pm$1.3 $\times10^6$ erg cm$^{-2}$.

Thus, neglecting the effects of other energy components and mass loss, such as gravitational potential energy and solar wind energy flux, the total energy introduced into the CH by the fast coronal wave was computed as  2.2$\pm$0.3 $\times10^8$ erg cm$^{-2}$ using the formula $\Delta E = \Delta E_t + \Delta E_r + \Delta E_c$.

When the fast coronal wave passes, the plasma within the CH likely experienced adiabatic compression and subsequent adiabatic expansion or rarefaction \citep{2018ApJ...864L..24L}.
To determine whether the observed temperature changes align with the expected temperature changes during the adiabatic process,
we have plotted temperature curves corresponding to the adiabatic process in Figure~5(b1)--(b3) as dashed lines.
These curves are based on the observed density changes represented by the blue curves in Figure~5(a1)--(a3) and the adiabatic equation $T/\rho^{\gamma-1} = T/n_e^{\gamma-1} = Constant$ \citep{2012ApJ...750..134D}, where $\rho$ denotes mass density, $\gamma$ is the specific heat ratio, equal to 5/3.
It is evident that the temperature curves observed in the three regions closely match the adiabatic temperature curves during the early stage of the rising phase.
This suggests that the plasma likely undergoes adiabatic compression during this period. However, it is also observed that the plasma temperature continues to rise after the initial adiabatic compression. 
This indicates that, in addition to adiabatic compression, wave dissipation and/or magnetic reconnection induced by the wave may have added more energy to the CH.
Furthermore, compared to the cooling of the adiabatic expansion, the observed temperature $T$ decreases more slowly after reaching their peaks, which may also reflect the influence of wave dissipation or reconnection heating.

After the plasma in the CH was adiabatically compressed by the wave, it might expand rapidly due to the magnetic tension, which would lead to a decrease in plasma temperature and thermal energy. 
If we exclude the effect of the adiabatic compression process, we can estimate the energy increase of the CH plasma that is due solely to wave dissipation, assuming the wave did not cause any other energy releases, such as reconnection heating.
According to our calculations, the increases in thermal energy, radiative energy loss, and thermal conduction energy resulting from adiabatic compression are approximately 1.8 $\times 10^8$ erg cm$^{-2}$, 3.5 $\times 10^6$ erg cm$^{-2}$, and 3.4 $\times 10^6$ erg cm$^{-2}$, respectively. Discounting these energies, the energy dissipated into the CH by the wave amounts to about 3.3 $\times 10^7$ erg cm$^{-2}$, which is only 15\% of the total increased energy $\Delta E$ of 2.2$\times10^8$ erg cm$^{-2}$.

When waves reach the boundary of  CHs, they are often observed to be reflected or undergo mode conversion \citep{2009ApJ...691L.123G, chen2016can, 2022A&A...659A.164Z}. The kinetic energy flux of the wave can be estimated by the formula $F_k = \rho (\delta v)^2 v_{ph}/2$ \citep{aschwanden2004role, liu2011direct}, where $v_{ph}$ is the phase speed of the wave, and $\delta v$ is the velocity perturbation. Assuming $\delta v/v_{ph} \geq \delta \rho/\rho$, it follows that $F_k \geq \frac{{\delta \rho}^2}{2\rho} {v_{ph}}^3 = \frac{(\delta n_e)^2 m_p {v_{ph}}^3}{2n_e}$, where $m_p$ denotes the proton mass. Our calculations show that the $F_k$ peaks within the CH, at the CH's boundary and in the primary region of the CH (excluding the boundary) with values of approximately 10.3, 22.7, and 5.2 $\times 10^5$ erg s$^{-1}$ cm$^{-2}$, which correspond to the kinetic energies per unit area $E_k$ of 2.2, 3.2, and 1.1 $\times 10^8$ erg cm$^{-2}$ over rising periods of  $\sim$7 minutes, respectively. This suggests that most ($\sim$70\%) of the wave energy is carried away by the reflected wave or converted into other forms of energy at the CH boundary.

\section{Discussions and Conclusions} \label{sec:dis}

In conclusion, this study effectively elucidated how a CH's characteristics were altered by a fast coronal wave, observed in both EUV and Lyman-alpha observations. The investigated coronal waves were associated with an X5.0 flare-related fast CME and comprised two distinct components. One component traveled swiftly across the CH at $\sim$950 km s$^{-1}$, which is comparable to the typical coronal Alfv\'{e}n speed ($\sim$1000 km s$^{-1}$), while the other, slower component moved at $\sim$470 km s$^{-1}$, exceeding the acoustic speed in the quiet corona ($\sim$230 km s$^{-1}$) and formed a stationary-mode wave that accumulated at the CH's boundary. 
This suggests that the two components may correspond to the shocked and fast-mode parts of the coronal magnetosonic wave, respectively.

We quantitatively examined the energy transferred from the wave to the CH through interactions, probably including adiabatic compression, wave-mode conversion, wave dissipation, and/or reconnection heating caused by the wave \citep{2014SoPh..289.3233L, 2018ApJ...864L..24L}.
Without factoring in extra energy and mass losses caused by solar wind transport, 
the coronal wave can deliver an energy per unit area of approximately $2.2 \times10^8$ erg cm$^{-2}$, including an increased radiation and conduction losses totaling roughly 7.8 $\times 10^6$ erg cm$^{-2}$, into the CH. 
With a rising duration of approximately 430 seconds, the average energy flux originating from the fast wave to the CH amounts to about $5.1 \times10^5$ erg s$^{-1}$ cm$^{-2}$, a value close to the typical solar wind energy flux of $7\times 10^5$ erg s$^{-1}$ cm$^{-2}$ \citep{withbroe1977mass} or the coronal heating requirement of $5 \times 10^5$ erg s$^{-1}$ cm$^{-2}$ in the CH \citep{1988ApJ...325..442W}. This outcome suggests that fast waves in the corona could significantly contribute to energizing solar winds or heating the corona in CHs at least during a short period of $\sim$7 mintues.
It is worth noting that if adiabatic expansion or rarefaction after compression is considered, the energy that the wave finally dissipates into the CH is $3.3 \times 10^7$ erg cm$^{-2}$, which is only about 15\% of the total increased energy of the CH.
Considering the changes in the $F_k$ peak or $E_k$ of the wave from the CH boundary to the CH main region, i.e. from $\sim$22.7 to 5.2 $\times 10^5$ erg s$^{-1}$ cm$^{-2}$, or from $\sim$3.2 to 1.1 $\times 10^8$ erg cm$^{-2}$, it is estimated that $\sim$70\% of the wave energy is carried away by the reflected wave or converted into other forms of energy at the CH boundary.

The CH ultimately reached a higher temperature but a decreased number density in a new equilibrium state about 1 hour after the wave passage.
Regarding the sharply reduced density, it is postulated that the input energy might propel a greater amount of plasma to move away from the CH, leading to a lower plasma density. 
The values of $T$ and $F_c$ exhibit non-negligible errors for the observed increases, as indicated by the 80\% of the Monte Carlo simulation results derived from the DEM method.
To determine whether this signal reflects a natural phenomenon, further high-resolution observations with enhanced spatial and temporal precision are necessary to validate the observed temperature elevation.
Additionally, we consider the scenario under the assumption that the mass and energy injected from the chromosphere into the CH remain constant before and after the wave's passage,
the slight temperature increase in the CH appears reasonable, because the radiative loss flux decreased sharply with $n_e$, while the thermal conduction flux did not change significantly (see the blue and green curves in Fig.~5(c1)--(c3)).
This study provides novel insights into comprehending the substantial energy influx within the CH, the alterations in CH properties, and the possible acceleration of solar winds following an eruption-driven fast wave.

\begin{acknowledgments}
This work is supported by the Strategic Priority Research Program of the Chinese Academy of Sciences, CAS (Grant No.XDB0560000), the National Key R\&D Program of China 2021YFA1600502, the National Natural Science Foundation of China (No. 12350004), Shenzhen Key Laboratory Launching Project (No. ZDSYS20210702140800001) and the Specialized Research Fund for State Key Laboratory of Solar Activity and Space Weather. H.D.C. was also supported by the Chinese Academy of Sciences (CAS) Scholarship. The SDO data are courtesy of NASA, the SDO/AIA, and SDO/HMI science teams. ASO-S mission is supported by the Strategic Priority Research Program on Space Science, the Chinese Academy of Sciences, Grant No. XDA15320000. CHIANTI is a collaborative project involving the University of Cambridge (UK), the NASA Goddard Space Flight Center (USA), the George Mason University (GMU, USA) and the University of Michigan (USA).
\end{acknowledgments}


\begin{thebibliography}{}
\bibitem[Aschwanden(2004)]{aschwanden2004role} Aschwanden, M.~J.\ 2004, SOHO 15 Coronal Heating, 575, 97
\bibitem[Chen et al.(2002)]{chen2002evidence} Chen, P.~F., Wu, S.~T., Shibata, K., et al.\ 2002, \apjl, 572, L99
\bibitem[Chen(2011)]{chen2011coronal} Chen, P.~F.\ 2011, LRSP, 8, 1
\bibitem[Chen et al.(2016)]{chen2016can} Chen, P.~F., Fang, C., Chandra, R., et al.\ 2016, \solphys, 291, 3195
\bibitem[Chen et al.(2019)]{chen2019lyman} Chen, B., Li, H., Song, K.-F., et al.\ 2019, RAA, 19, 159
\bibitem[Chen et al.(2024)]{chen2024simultaneous} Chen, H., Fletcher, L., Zhou, G., et al.\ 2024, \apj, 976, 207
\bibitem[Cheung et al.(2015)]{cheung2015thermal} Cheung, M.~C.~M., Boerner, P., Schrijver, C.~J., et al.\ 2015, \apj, 807, 143
\bibitem[Del Zanna et al.(2021)]{del2021chianti} Del Zanna, G., Dere, K.~P., Young, P.~R., et al.\ 2021, \apj, 909, 38
\bibitem[Downs et al.(2012)]{2012ApJ...750..134D} Downs, C., Roussev, I.~I., van der Holst, B., et al.\ 2012, \apj, 750, 2, 134
\bibitem[Feng et al,(2019)]{feng2019lyman} Feng, L., Li, H., Chen, B., et al.\ 2019, RAA, 19, 162
\bibitem[Fletcher et al.(2011)]{fletcher2011observational} Fletcher, L., Dennis, B.~R., Hudson, H.~S., et al.\ 2011, \ssr, 159, 19
\bibitem[Gallagher et al.(1999)]{gallagher1999radial} Gallagher, P.~T., Mathioudakis, M., Keenan, F.~P., et al.\ 1999, \apjl, 524, L133
\bibitem[Gan et al.(2019)]{gan2019advanced} Gan, W.-Q., Zhu, C., Deng, Y.-Y., et al.\ 2019, RAA, 19, 156
\bibitem[Gopalswamy et al.(2009)]{2009ApJ...691L.123G} Gopalswamy, N., Yashiro, S., Temmer, M., et al.\ 2009, \apjl, 691, 2, L123
\bibitem[Hannah \& Kontar(2012)]{2012A&A...539A.146H} Hannah, I.~G. \& Kontar, E.~P.\ 2012, \aap, 539, A146
\bibitem[Hou et al.(2022)]{hou2022three} Hou, Z., Tian, H., Wang, J.-S., et al.\ 2022, \apj, 928, 98
\bibitem[Huber et al.(1974)]{huber1974extreme} Huber, M.~C.~E., Foukal, P.~V., Noyes, R.~W., et al.\ 1974, \apjl, 194, L115
\bibitem[Kienreich et al.(2011)]{2011ApJ...727L..43K} Kienreich, I.~W., Veronig, A.~M., Muhr, N., et al.\ 2011, \apjl, 727, L43
\bibitem[Kozarev et al.(2011)]{2011ApJ...733L..25K} Kozarev, K.~A., Korreck, K.~E., Lobzin, V.~V., et al.\ 2011, \apjl, 733, L25
\bibitem[Lemen et al.(2012)]{lemen2012atmospheric} Lemen, J.~R., Title, A.~M., Akin, D.~J., et al.\ 2012, \solphys, 275, 17
\bibitem[Li et al.(2019)]{li2019lyman} Li, H., Chen, B., Feng, L., et al.\ 2019, RAA, 19, 158
\bibitem[Li et al.(2019)]{2019RAA....19..165L} Li, C., Fang, C., Li, Z., et al.\ 2019, RAA 19, 165
\bibitem[Li et al.(2022)]{2022SCPMA..6589602L} Li, C., Fang, C., Li, Z., et al.\ 2022, SCPMA, 65, 289602
\bibitem[Liu et al. (2011)]{liu2011direct} Liu, W., Title, A.~M., Zhao, J., et al.\ 2011, \apjl, 736, L13
\bibitem[Liu et al.(2012)]{liu2012quasi} Liu, W., Ofman, L., Nitta, N.~V., et al.\ 2012, \apj, 753, 52
\bibitem[Liu \& Ofman(2014)]{2014SoPh..289.3233L} Liu, W. \& Ofman, L.\ 2014, \solphys,  289, 9, 3233.
\bibitem[Liu et al.(2018)]{2018ApJ...864L..24L} Liu, W., Jin, M., Downs, C., et al.\ 2018, \apjl, 864, 2, L24. 
\bibitem[Liu et al.(2022)]{2022SCPMA..6589605L} Liu, Q., Tao, H., Chen, C., et al.\ 2022, SCPMA, 65, 289605
\bibitem[Long et al.(2008)]{long2008kinematics} Long, D.~M., Gallagher, P.~T., McAteer, R.~T.~J., et al.\ 2008, \apjl, 680, L81
\bibitem[Munro \& Withbroe(1972)]{munro1972properties} Munro, R.~H. \& Withbroe, G.~L.\ 1972, \apj, 176, 511
\bibitem[Nitta et al.(2013)]{2013ApJ...776...58N} Nitta, N.~V., Schrijver, C.~J., Title, A.~M., et al.\ 2013, \apj, 776, 58
\bibitem[Olmedo et al.(2012)]{olmedo2012secondary} Olmedo, O., Vourlidas, A., Zhang, J., et al.\ 2012, \apj, 756, 143
\bibitem[Pesnell et al.(2012)]{pesnell2012solar} Pesnell, W.~D., Thompson, B.~J., \& Chamberlin, P.~C.\ 2012, \solphys, 275, 3
\bibitem[Piantschitsch et al.(2018)]{2018ApJ...860...24P} Piantschitsch, I., Vr{\v{s}}nak, B., Hanslmeier, A., et al.\ 2018, \apj, 860, 1, 24.
\bibitem[Piantschitsch et al.(2024)]{2024A&A...687A.200P} Piantschitsch, I., Terradas, J., Soubrie, E., et al.\ 2024, \aap, 687, A200.
\bibitem[Priest(2014)]{priest2014magnetohydrodynamics} Priest, E.\ 2014, Magnetohydrodynamics of the Sun, by Eric Priest, Cambridge, UK: Cambridge University Press, 2014
\bibitem[Qiu et al.(2022)]{2022SCPMA..6589603Q} Qiu, Y., Rao, S., Li, C., et al.\ 2022, SCPMA, 65, 289603
\bibitem[Rosner \& Vaiana(1977)]{rosner1977hydrostatic} Rosner, R. \& Vaiana, G.~S.\ 1977, \apj, 216, 141
\bibitem[Ryan et al.(2024)]{ryan2024triangulation} Ryan, D.~F., Massa, P., Battaglia, A.~F., et al.\ 2024, \solphys, 299, 114
\bibitem[Scherrer et al.(2012)]{scherrer2012helioseismic} Scherrer, P.~H., Schou, J., Bush, R.~I., et al.\ 2012, \solphys, 275, 207
\bibitem[Schmidt \& Ofman(2010)]{2010ApJ...713.1008S} Schmidt, J.~M. \& Ofman, L.\ 2010, \apj, 713, 2, 1008. 
\bibitem[Su et al.(2013)]{su2013imaging} Su, Y., Veronig, A.~M., Holman, G.~D., et al.\ 2013, NatPh, 9, 489
\bibitem[Su et al.(2019)]{su2019simulations} Su, Y., Liu, W., Li, Y.-P., et al.\ 2019, RAA, 19, 163
\bibitem[Vanninathan et al.(2015)]{2015ApJ...812..173V} Vanninathan, K., Veronig, A.~M., Dissauer, K., et al.\ 2015, \apj, 812, 173
\bibitem[Veronig et al.(2008)]{veronig2008high} Veronig, A.~M., Temmer, M., \& Vr{\v{s}}nak, B.\ 2008, \apjl, 681, L113
\bibitem[Veronig et al.(2018)]{2018ApJ...868..107V} Veronig, A.~M., Podladchikova, T., Dissauer, K., et al.\ 2018, \apj, 868, 107
\bibitem[Wang et al.(2021)]{wang2021exploring} Wang, C., Chen, F., \& Ding, M.\ 2021, \apjl, 911, L8
\bibitem[Warmuth(2015)]{warmuth2015large} Warmuth, A.\ 2015, LRSP, 12, 3
\bibitem[Withbroe \& Noyes(1977)]{withbroe1977mass} Withbroe, G.~L. \& Noyes, R.~W.\ 1977, \araa, 15, 363
\bibitem[Withbroe(1988)]{1988ApJ...325..442W} Withbroe, G.~L.\ 1988, \apj, 325, 442
\bibitem[Zhang et al.(2019)]{zhang2019hard} Zhang, Z., Chen, D.-Y., Wu, J., et al.\ 2019, RAA, 19, 160
\bibitem[Zhang et al.(2019)]{zhang2019energy} Zhang, Q.~M., Cheng, J.~X., Feng, L., et al.\ 2019, \apj, 883, 124
\bibitem[Zheng et al.(2018)]{zheng2018extreme} Zheng, R., Chen, Y., Feng, S., et al.\ 2018, \apjl, 858, L1
\bibitem[Zheng(2024)]{zheng2024recent} Zheng, R.\ 2024, RSPSA, 480, 20230950
\bibitem[Zhou et al.(2020)]{2020ApJ...905..150Z} Zhou, G., Gao, G., Wang, J., et al.\ 2020, \apj, 905, 150
\bibitem[Zhou et al.(2022)]{2022A&A...659A.164Z} Zhou, X., Shen, Y., Tang, Z., et al.\ 2022, \aap, 659, A164. 
\bibitem[Zhou et al.(2024)]{zhou2024resolved} Zhou, X., Shen, Y., Yuan, D., et al.\ 2024, NatCo, 15, 3281
\end{thebibliography}



\end{document}